\documentclass[11pt]{article}
\usepackage{graphicx}

\newcommand{\BABARPubYear}    {05}

\newcommand{\BABARConfNumber} {14}
\newcommand{\SLACPubNumber} {11368}

\input babarsym

\newcommand{\psfile}[3][]{ 
  \begin{center}
    \setlength{\epsfxsize}{#3\linewidth}\leavevmode
    \def\noOpt{}\def\testit{#1}\ifx\testit\noOpt%
      \epsfbox{#2}%
    \else%
      \epsfbox[#1]{#2}%
    \fi
  \end{center}
}

\providecommand{\xf}{\mbox{${\cal F}$}}

\providecommand{\DE}{\ensuremath{\Delta E}}

\providecommand{\half}{\mbox{${1\over2}$}}

\providecommand{\etapr}{\mbox{$\eta^{\prime} $}}

\providecommand{\etapKz}{\mbox{$\eta^{\prime} K^0$}}
\providecommand{\etapKzs}{\mbox{$\eta^{\prime} K^0_S$}}
\providecommand{\etapKzl}{\mbox{$\eta^{\prime} K^0_L$}}

\def\deltaS{\ensuremath{{\rm \Delta}S}\xspace}

\newcommand{\fetapreppkl}{\ensuremath{\etapr_{\eta\pi\pi} K_{L}^{0}}}
\newcommand{\fetaprrgkl}{\ensuremath{\etapr_{\rho\gamma} K_{L}^{0}}}
\newcommand{\fetapreppggkz}{\ensuremath{\etapr_{\eta(\gamma\gamma)\pi\pi} K^{0} }}
\newcommand{\fetapreppthrpikz}{\ensuremath{\etapr_{\eta(3\pi)\pi\pi}K^{0}}}
\newcommand{\hel}{\mbox{${\cal H}$}}
\newcommand{\msp}{\phantom{-}}

\providecommand{\EtapEtaPiPi}{\mbox{$\etapr \rightarrow \eta \pi^+ \pi^-$}}
\providecommand{\EtapRhoGam}{\mbox{$\etapr \rightarrow \rho^0  \gamma$}}

\providecommand{\fEtapEtaPiPi}{\mbox{$\etapr_{\eta \pi \pi}$}}
\providecommand{\fEtapRhoGam}{\mbox{$\etapr_{\rho\gamma}$}}

\providecommand{\EtaGG}{\mbox{$\eta \rightarrow \gamma  \gamma$}}

\providecommand{\mgg}{\mbox{$m_{\gamma \gamma}$}}

\providecommand{\mpipi}{\mbox{$m_{\pi \pi}$}}

\providecommand{\fetaprgKz}{\mbox{$\etapr_{\rho\gamma} K^0$}}

\providecommand{\tcp}{\mbox{$t_{CP}$}}
\providecommand{\ttag}{\mbox{$t_{\rm tag}$}}

\providecommand{\BetapKz}{\mbox{$B^0 \rightarrow \eta^{\prime} K^0$}}

\providecommand{\BetapKzl}{\mbox{$B^0 \rightarrow \eta^{\prime} K^0_L$}}

\providecommand{\UfourS}{\mbox{$\Upsilon(4S)$}}
\providecommand{\epem}{\mbox{$e^+e^-$}}
\def\BB{\mbox{$B\overline B\ $}}
\def\pep2{PEP-II}

\providecommand\etal{{\it et al.}}

\long\def\inst#1{\par\nobreak\kern 4pt\nobreak
    {\it #1}\par\vskip 10pt plus 3pt minus 3pt}

\begin{document}
{\pagestyle{empty}

\begin{flushright}
\babar-CONF-\BABARPubYear/\BABARConfNumber \\
SLAC-PUB-\SLACPubNumber \\
\end{flushright}

\par\vskip 3cm

\begin{center}
\Large \bf Measurement of Time-Dependent 
{\boldmath \CP }-Violating Asymmetries in 
{\boldmath \Bz }  Meson Decays to {\boldmath \etapKzl }
\end{center}
\bigskip

\begin{center}
\large The \babar\ Collaboration\\
\mbox{ }\\
\today
\end{center}
\bigskip \bigskip

\begin{center}
\large \bf Abstract
\end{center}

We present a  preliminary measurement of \CP -violating parameters $S$ and $C$ 
from fits of the time-dependence of \Bz\ meson decays to \etapKzl . The data 
were recorded with the \babar\ detector at \pep2\ and correspond to 
$232  \times 10^6$ \BB\ pairs produced in \epem\ annihilation through the
\UfourS\ resonance. By fitting the time-dependent \CP\ asymmetry of the 
reconstructed $\Bz \to \etapKzl$ events, we find $S = 0.60 \pm 0.31 \pm 0.04$
 and 
$C = 0.10 \pm 0.21 \pm 0.03$, where the first error quoted is statistical 
and the second is systematic.
We also perform a combined fit using both \etapKzs\ and \etapKzl\ data, and
find $S = 0.36 \pm 0.13  \pm 0.03$  and 
$C = -0.16 \pm 0.09 \pm 0.02$.

\vfill
\begin{center}
Submitted at the 
International Europhysics Conference On High-Energy Physics (HEP 2005),
7/21---7/27/2005, Lisbon, Portugal
\end{center}

\vspace{1.0cm}
\begin{center}
{\em Stanford Linear Accelerator Center, Stanford University, 
Stanford, CA 94309} \\ \vspace{0.1cm}\hrule\vspace{0.1cm}
Work supported in part by Department of Energy contract DE-AC03-76SF00515.
\end{center}

\newpage
} 

\begin{center}
\small

The \babar\ Collaboration,
\bigskip

B.~Aubert,
R.~Barate,
D.~Boutigny,
F.~Couderc,
Y.~Karyotakis,
J.~P.~Lees,
V.~Poireau,
V.~Tisserand,
A.~Zghiche
\inst{Laboratoire de Physique des Particules, F-74941 Annecy-le-Vieux, France }
E.~Grauges
\inst{IFAE, Universitat Autonoma de Barcelona, E-08193 Bellaterra, Barcelona, Spain }
% ...

\end{center}\newpage

\section{Introduction}
\label{sec:Introduction}

Decays of $B^0$ mesons to charmless hadronic final states such as
$\phi K^0$, $K^+ K^- K^0$, $\etapr K^0$, $\piz K^0$ and $f_0(980) K^0$
proceed mostly via a single penguin (loop) amplitude with the same weak
phase as in  $B$ decays to a charmonium state plus a $K^0$ meson \cite{Penguin,soni}.
However Cabibbo-Kobayashi-Maskawa (CKM)-suppressed amplitudes and new  particles in
the loop can introduce other weak phases whose contribution is not negligible \cite{Penguin,lonsoni}.
 
\begin{figure}[htbp]
\begin{center}
\vspace{4cm}
\includegraphics[bb=85 155 535 605,angle=0,scale=0.8]{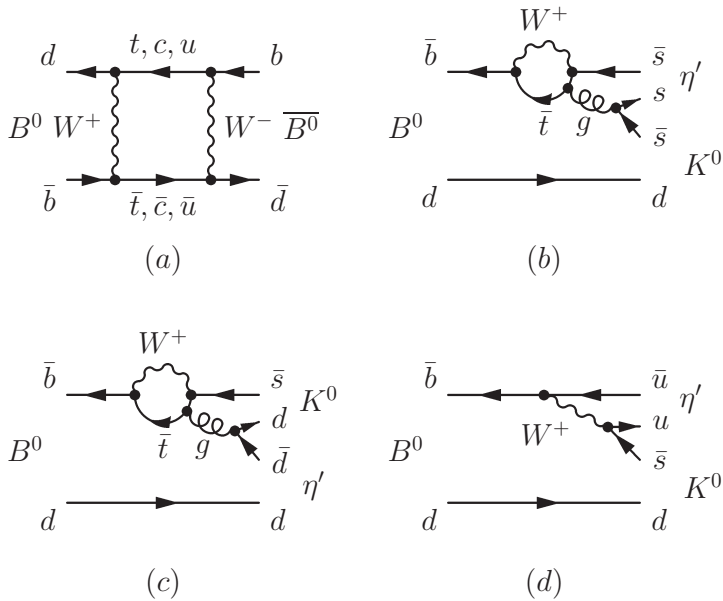}
\vspace{-9cm}
\caption{Feynman diagrams describing (a) $B-\bar{B}$ mixing, the decay
 $B^0 \rightarrow \eta^{\prime} K^0$ via (b, c) internal gluonic penguin and  
(d) color-suppressed tree. }
\end{center}
  \label{fig:Feyn}
\end{figure}

Fig.~\ref{fig:Feyn}(a) shows the diagram describing the 
$B-\bar{B}$ mixing.
The amplitudes shown in  Fig.~\ref{fig:Feyn}(b)-(d) are relevant for the
decay \BetapKz. 
All of the amplitudes are suppressed by small CKM 
matrix elements, but the tree 
diagram for $B^0$ shown in Fig.~\ref{fig:Feyn}(d) is expected
to be smaller \cite{soni,beneke} since there is additional CKM suppression
and color suppression.

For the decay \BetapKz, the additional contributions of other weak phases 
from CKM-suppressed amplitudes are expected to be small,
so the time-dependent \CP\ asymmetry measurement  provides an
approximate measurement of \stwob. We define $S$ and $C$ as the coefficients 
of sine and cosine oscillation terms, respectively, in the $B \bar{B}$ decay
 rate distributions (see Eqn.~(\ref{fplusminus}) below).
\deltaS is the  deviation between $S$ 
in the decay \BetapKz\
and $S$, equal to \stwob ,   in the charmonium $K^0$
decays.  Theoretical bounds for this deviation have been calculated with
an SU(3) analysis \cite {Gross,Gronau}. Such bounds have been improved
 \cite{Gronau2} by
the  measurements of \Bz\ decays to a pair of neutral charmless light
pseudoscalar mesons \cite{Isosca,PRD}, with the
conclusion that \deltaS\ is expected to be less than 0.10 (with a theoretical
uncertainty less than $\sim$0.03 due to the assumptions in the calculation).
QCD factorization  calculations conclude that \deltaS\ is even smaller \cite{BN}.
A significantly larger value of \deltaS\ could arise from non-Standard-Model
amplitudes \cite{lonsoni}.

The  \CP -violating asymmetry in the decay \BetapKz\  has been  measured 
previously by the \babar\ ~\cite{Previous}
and Belle ~\cite{BELLE} experiments using the final state  \etapKzs . 
In the measurement presented  in this paper we use events reconstructed 
in the \etapKzl\ final state.
 We  present also the measurement  of  \CP -violating asymmetry  
combining present \etapKzl\ data  with the  \etapKzs\ data  used in the 
previous \babar\  measurement \cite{Previous}.
\section{The \babar\ Detector and Dataset}
\label{sec:babar}

The results presented in this paper are based on data collected
in 1999--2004 with the \babar\ detector~\cite{babar2}
at the PEP-II asymmetric $e^+e^-$ collider~\cite{pep}
located at the Stanford Linear Accelerator Center.  An integrated
luminosity of 211~fb$^{-1}$, corresponding to about 
232  million \BB\ pairs, was recorded at the $\Upsilon (4S)$ resonance
(``on-resonance'', center-of-mass energy $\sqrt{s}=10.58\ \gev$).

The asymmetric beam configuration in the laboratory frame
provides a boost of $\beta\gamma = 0.56$ to the $\Upsilon(4S)$.
Charged particles are detected and their momenta measured by the
combination of a silicon vertex tracker (SVT), consisting of five layers
of double-sided detectors, and a 40-layer central drift chamber,
both operating in the 1.5 T magnetic field of a solenoid.
The tracking system covers 92\% of the solid angle in the center-of-mass (CM) frame.

Charged-particle identification (PID) is provided by the average
energy loss (\dedx) in the tracking devices  and
by an internally reflecting ring-imaging
Cherenkov detector (DIRC) covering the central region.
A $K/\pi$ separation of better than four standard deviations ($\sigma$)
is achieved for momenta below 3 \gevc , decreasing to 2.5 $\sigma$ at the highest
momenta in the $B$ decay final states.
Photons and electrons are detected by a CsI(Tl) electromagnetic calorimeter (EMC).
The EMC provides good energy and angular resolutions for detection of photons in
 the range from 30 \mev\ to 4 \gev. The energy and angular resolutions are 3\% 
and 4 \mrad, respectively, for a 1 \gev\ photon.

The flux return (IFR) for the solenoid  is composed of multiple layers of iron
and resistive plate chambers for the identification of muons and long-lived
neutral hadrons.

\section{Event Selection and Analysis Method}
Monte Carlo (MC) simulations  of the signal decay modes, \BB\ 
backgrounds, and detector response are used to establish the event selection criteria.
We reconstruct 
\etapr\ mesons  through the decays  \EtapRhoGam\ (\fEtapRhoGam ) 
and   \EtapEtaPiPi\ (\fEtapEtaPiPi )  with \EtaGG .
The photon energy $E_{\gamma}$ in laboratory system 
must be greater than 50 \mev\ for $\eta$ candidates,
and greater than 200 \mev\ in  \fEtapRhoGam . We make the following 
requirements on the invariant masses (in \mevcc ): 490 $<$ \mgg $<$ 600  for 
$\eta$, $510 < \mpipi\ < 1000$ for $\rho^0$,
$945< m_{\eta\pi\pi} < 970$ for \fEtapEtaPiPi , and $930<m_{\rho\gamma} < 980$ for \fEtapRhoGam.
We require the PID information of the signal pions to be consistent with the 
pion hypothesis.

Signal \KL\ candidates are reconstructed from clusters of energy deposited in the EMC 
or from hits in the IFR  
not associated to any charged track in the event. Because the energy of the \KL\ is not measured,
 we determine the \KL\ candidate  laboratory momentum from  its flight direction determined from the \etapr\ vertex and 
the centroid of the EMC (or IFR) candidate 
 and the constraints of \KL\ and $B^0$ masses to their nominal 
values~\cite{PDG2004}.

A $B$ meson candidate is characterized kinematically by the 
energy difference $\DE \equiv E_B^*-\half\sqrt{s}$, where $E_B^*$ is the
 CM energy of the $B$ meson. Signal events are peaked within
$\pm10$ \mev of $\DE = 0$ while background events extend towards 
positive values of \DE\
(this is a consequence of the mass constraint used to
determine the \KL\ momentum). We require $-0.01 < \DE < 0.08$ \gev . 
This choice is dictated by the need of preserving a region
with enough background  for  a fit to that component.
 
To reject  background due to continuum $\epem\ra\qqbar$ events ($q=u,d,s,c$),
 we make use
of the angle $\theta_T$ between the thrust axis of the $B$ candidate and
that of
the rest of the tracks and neutral clusters in the event, calculated in
the center-of-mass frame.  The distribution of $\cos{\theta_T}$ is
sharply peaked near $\pm1$ for combinations drawn from jet-like $q\bar q$
pairs and is nearly uniform for the isotropic $B$ meson decays; we require
$|\cos{\theta_T}|<0.8 $ in the \EtapEtaPiPi\ subdecay mode and
 $|\cos{\theta_T}|<0.75 $ in the \EtapRhoGam\ subdecay.
 
For further suppression of continuum background we require  
that the total missing transverse momentum projected
along the \KL\ direction, where the total momentum is calculated with all
charged tracks and neutral clusters (without the \KL ), is
no more than 0.45 \gevc lower than the calculated  
transverse momentum of the \KL\ candidate.
We also require that the 
cosine of the  polar angle $\theta$ of the total missing 
momentum in laboratory system to be less than 0.95.

The purity of the \KL\ candidates reconstructed in the EMC is further improved 
by a cut on the output of a neural  network (NN) that takes cluster-shape
 variables as its inputs. The NN was trained using MC signal events 
and data events in the side band
distribution (defined as \mbox{$0.04<\DE<0.08$ \gev }).
We validated the performance of the NN using \KL\ candidates in
the reconstructed $\Bz \to J/\psi K^0_L$ events.

All selection criteria have been chosen using MC signal and background events
to maximize the expected statistical significance of signal yield in the data.

The \BB\ backgrounds were estimated using Monte Carlo simulations 
of \BzBzb\ and \BpBm . 
We found a small evidence of \BB\ background from $b\ra c$ decays  in the sub-decay 
mode   \EtapRhoGam\ , so we added this  component to the fit.

For each $\Bz \to \etapKzl$ candidate ($B_{CP}$), we reconstruct the decay 
vertex of the other $B$ meson  ($B_{\rm tag}$) from the remaining 
charged tracks in the event and identify its flavor.
The time difference $\deltat \equiv \tcp - \ttag$,
where $\tcp$ and $\ttag$ are the proper decay times of the $B_{CP}$ and 
$B_{\rm tag}$, respectively, is obtained from the measured distance between the $B_{CP}$
and  $B_{\rm tag}$ decay vertices and from the boost ($\beta \gamma =0.56$) of 
the \epem beam system. The distribution of \deltat\ is:
\begin{equation}
F( {\deltat}) = \frac{e^{-\left|\deltat\right|/\tau}}{4\tau} \left\{1 \mp {\Delta \omega} \pm (1 -2 \omega) \left[-\eta S \sin(\deltamd
\deltat) - C \cos(\deltamd\deltat)\right]\right\},
\label{fplusminus}
\end{equation}
where the upper (lower) sign denotes a decay accompanied by a \Bz\ (\Bzb)
 tag, $\tau$ is the mean \Bz\ lifetime, \deltamd\ is the mixing frequency,
$\eta$ is the \CP\ eigenvalue of the final state ($\eta=+1$ for \etapKzl , $\eta=-1$ for \etapKzs ) and the mistag parameters $\omega$ and $\Delta \omega$ are the average and difference, 
respectively, of the probabilities that 
a true \Bz\ (\Bzb ) meson is tagged as \Bzb\ (\Bz ). The tagging algorithm, based on six tagging categories, is an improved version of what was used in the  previous \babar\ publication
 \cite{Previous}. Separate neural networks are trained to identify 
primary leptons, kaons, soft pions from $D^*$ decays, and high-momentum 
charged particles from \B\ decays.
Each event is assigned to one of the tagging 
categories based on the  source of tagging information and on the   estimated 
mistag probability. The distribution $F(\deltat )$ is convolved 
with a resolution function to account for the finite vertex resolution of the detector.

\section{Maximum Likelihood Fit}
We use an unbinned, multivariate maximum-likelihood fit to extract
signal yields and \CP\-violating parameters.
We use the following discriminating variables: 
\DE , a Fisher discriminant \xf, \deltat . In the decay mode  \EtapRhoGam\ we add the   \etapr\ mass and the  variable  \hel , defined as the cosine 
of the $\rho$ meson's rest frame decay angle of a pion with respect to the $\eta^{\prime}$ 
flight direction.
The Fisher discriminant combines four variables: the angles with respect to the beam axis of the $B$
momentum and $B$ thrust axis in the \UfourS\ frame, and the zeroth and second
angular moments of the energy flow excluding the $B$ candidate around the
$B$ thrust axis.

We indicate with $j$ the event species:
 signal, continuum background, or \BB\ background.
For each species $j$ and each flavor-tagging category $c$, we define a total 
probability density function (PDF)
for an events $i$ as:
\begin{equation}
{\cal P}_{j,c}^i =  {\cal  P}_j ( \DE^i ) 
\, { \cal P}_j( \xf^i ) 
\, { \cal P}_j( M^i_{\etapr} ) 
\, { \cal P}_j( {\cal H}^i ) 
\, { \cal  P}_j (\Delta t^i , \sigma_{\Delta t}^i,c).
\end{equation}
where $\sigma_{\deltat}^i$ is the error on \deltat\ for an event $i$.  
We define the extended likelihood function for the $N_c$ input events in 
category $c$ as 
\begin{equation}
{\cal L}_c= \exp{\left(-\sum_j n_j f_{j,c}\right)}
           \prod_i^{N_c} (n_{\rm sig}f_{{\rm sig},c}{\cal P}_{{\rm sig},c}^{i}
                   +n_{q\bar{q}} f_{q\bar{q},c}{\cal P}_{q\bar{q}}^{i}
                   +n_{B\bar{B}}f_{B\bar{B},c}{\cal P}_{B\bar{B}}^{i}),
\end{equation}
where $n_j$ is the number of events with species $j$, and $f_{j,c}$ is the fraction 
of category-$c$ events with species $j$. 
We fix $f_{sig,c}$ and $f_{B \bar{B},c}$ to $f_{B_{flav},c}$, the values measured with a  
sample of neutral $B$ decays to flavor eigenstates, $B_{flav}$.

The total likelihood function for all categories is given as the
product of the likelihoods over the seven tagging categories (including a category for
 untagged events for yield determination).

\section{Results}
\label{sec:Physics}
The reconstruction efficiency  is  10.3\% and 11.6\% for
\fetaprrgkl\ and \fetapreppkl\ respectively. 
In Table~\ref{tab:NewMeas} we give the number of the signal yield and 
the parameters $S$ and $C$. Note that the sign of the \CP\ eigenvalue
of the final state is out of the definition of  $S$ parameter (see    
Eqn.~(\ref{fplusminus})), so  $S$ parameter has the same sign 
in  \etapKzs\ and \etapKzl\ events.   The \etapKzs\ data are those used in 
 \babar\  previous measurement \cite{Previous}.
Combining \fetapreppkl\ events
and  \fetaprrgkl\ events, we measure $S = 0.60 \pm 0.31$ and $C= 0.10 \pm 0.21$.
In this fit we have 42 free parameters: $S$, $C$, signal yields (2),
\BB\ background yield (1), continuum background yields (2) and fractions (12),
background \deltat , \DE , \xf , \etapr\ mass and \hel\ PDF parameters (23).
In the final fit, combining  \etapKzs\ and  \etapKzl ,  
we have 138 free parameters: $S$, $C$, signal yields (7),
\BB\ background yield (3), continuum background yields (7) and fractions (42),
background \deltat , \DE , \xf , \etapr\ mass and \hel\ PDF parameters (77).

\begin{table}[!hb]
\caption{Results with statistical errors for the $\Bz\to\etapr K^0$
time-dependent fits (decays with \KL\  in upper part on the table
and decays with \KS\  in lower part of the table).}
\label{tab:NewMeas}
\begin{center}
\vspace*{-0.3cm}
\begin{tabular}{lccc}
\hline\hline
Mode                     &Signal yield &       $S$       &         $C$      \\
\hline
$\fetapreppkl$  &  $137\pm22$  &   $\msp0.38\pm0.44$&$\msp0.34\pm0.29$ \\
$\fetaprrgkl$  &  $303\pm49$  &   $\msp0.88\pm0.43$&$-0.15\pm0.29$ \\
\hline
$\fetapreppggkz_{\pi^+\pi^-}$      & $188\pm15$  &$\msp 0.01\pm0.28$&   $-0.18\pm0.18$ \\
$\fetaprgKz_{\pi^+\pi^-}$       & $430\pm26$  &$\msp0.44\pm0.19$&   $-0.30\pm0.13$ \\
$\fetapreppthrpikz_{\pi^+\pi^-}$&$54\pm8$ &$\msp0.79\pm0.47$&$\msp0.11\pm0.35$ \\
$\fetapreppggkz_{\pi^{0}\pi^{0}}$      &$44\pm9$ &   $-0.04\pm0.57$&   $-0.65\pm0.42$ \\
$\fetaprgKz_{\pi^{0}\pi^{0}}$       &  $89\pm23$  &   $-0.45\pm0.68$&$\msp0.41\pm0.40$ \\
\hline

Combined fit                  & $1245\pm67$  &$\msp0.36\pm0.13$&   $-0.16\pm0.09$ \\
\hline\hline
\end{tabular}
\end{center}
\vspace*{-0.3cm}
\end{table}

\begin{figure}[tbp]
 \begin{minipage}{\linewidth}
  \begin{center}
  \includegraphics[bb=85 155 535 605,scale=0.35]{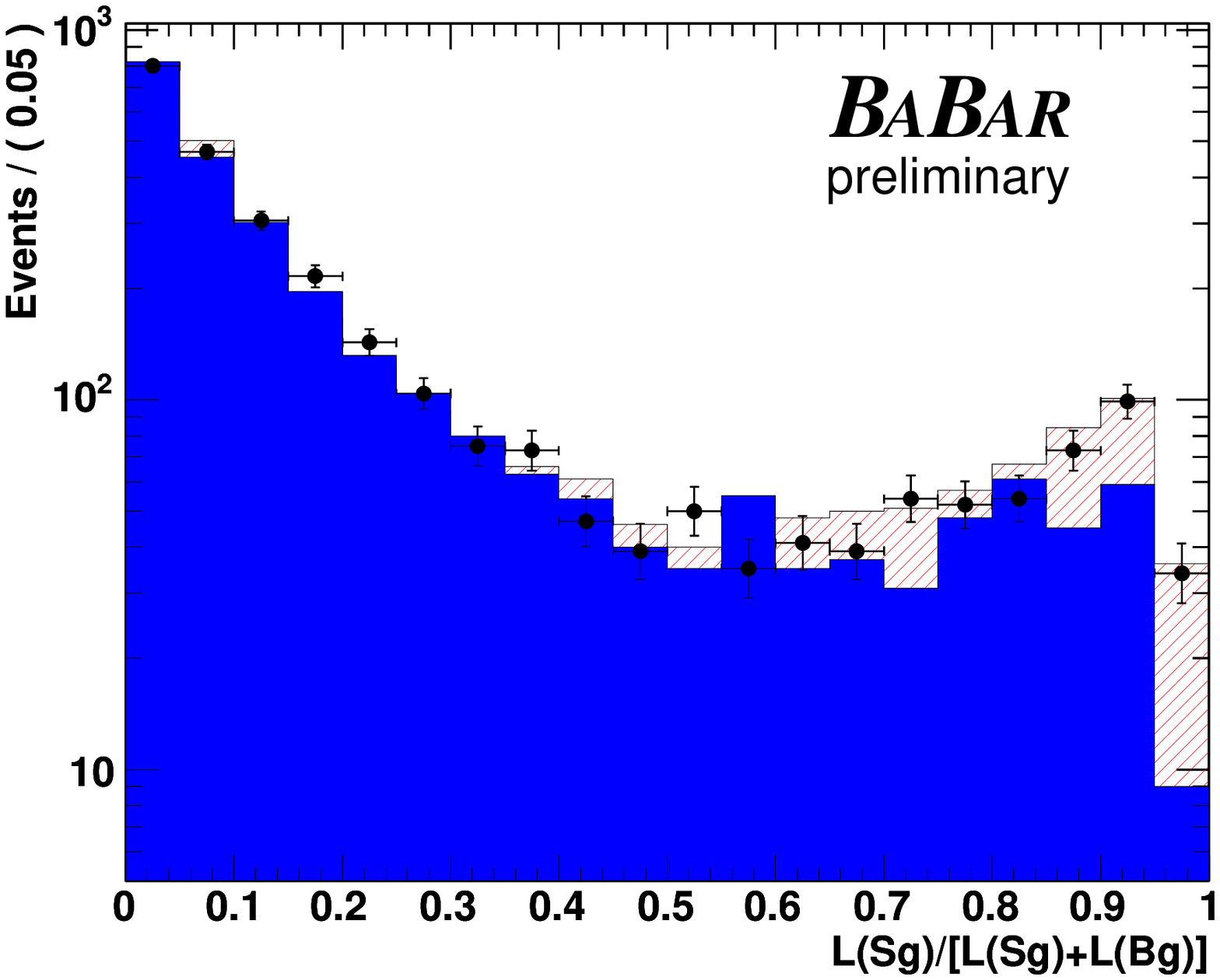}
  \hspace{2cm}
  \includegraphics[bb=85 155 535 605,scale=0.35]{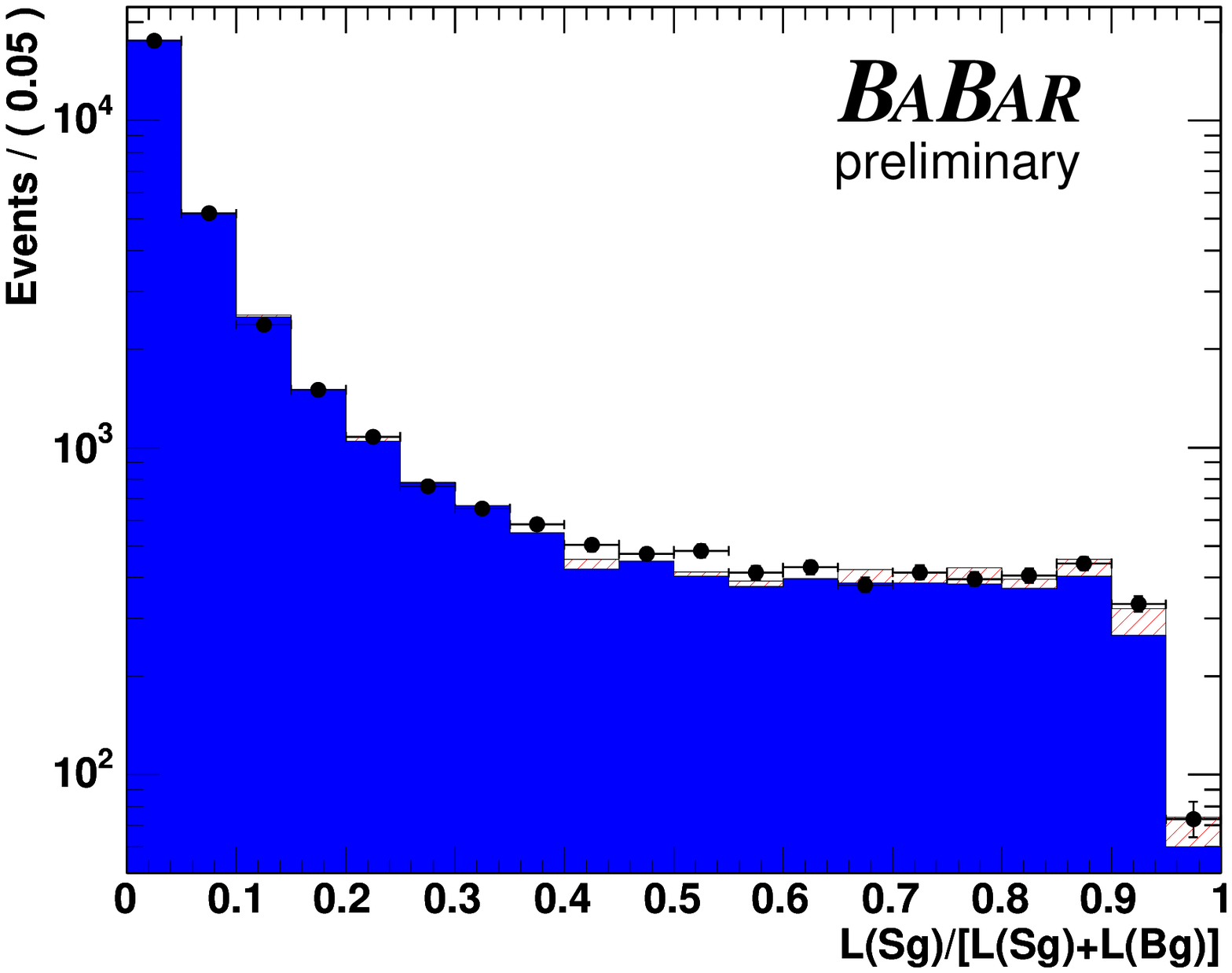}
   \end{center}
  \end{minipage}
  \vspace{2cm}
\caption{The likelihood ratio $\calL(Sg)/[\calL(Sg)+\calL(Bg)]$ for
\fetapreppkl\ (left) and \fetaprrgkl\ (right). The points represent the
on-resonance data, the histograms are from PDF generated events of
background (blue area) and background plus signal (shaded red area). 
}
\label{fig:ProbRatioDistri}
\end{figure}

\begin{figure}[!htbp]
\begin{minipage}{\linewidth}
\begin{center}
\includegraphics[scale=0.5]{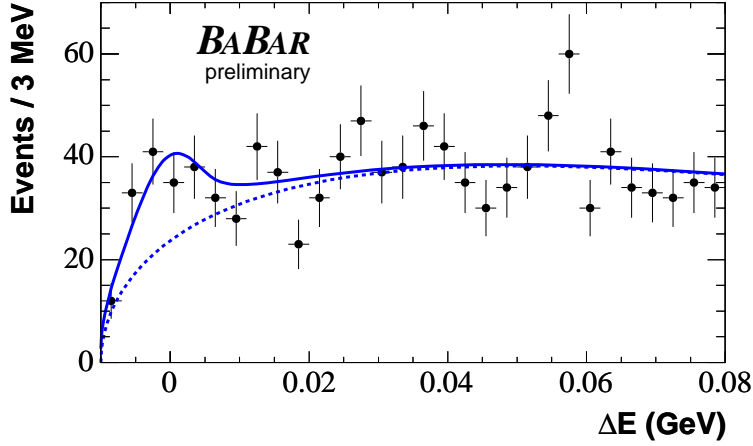}
\end{center}
\end{minipage}
\caption{Projection onto \DE\ for $\Bz \to \etapr \KL$ 
(sum of the sub-decay modes \fetaprrgkl\ and \fetapreppkl )
of a subset of the data for which the signal likelihood 
(computed without the plotted variable) exceeds a mode-dependent threshold 
that optimizes the sensitivity.
Points with errors represent the data,
solid curve the full fit functions and dashed blue curve the total 
background functions. }
\label{fig:DEplot}
\end{figure}

\begin{figure}[!htb]
\begin{center}
   \includegraphics[scale=0.5]{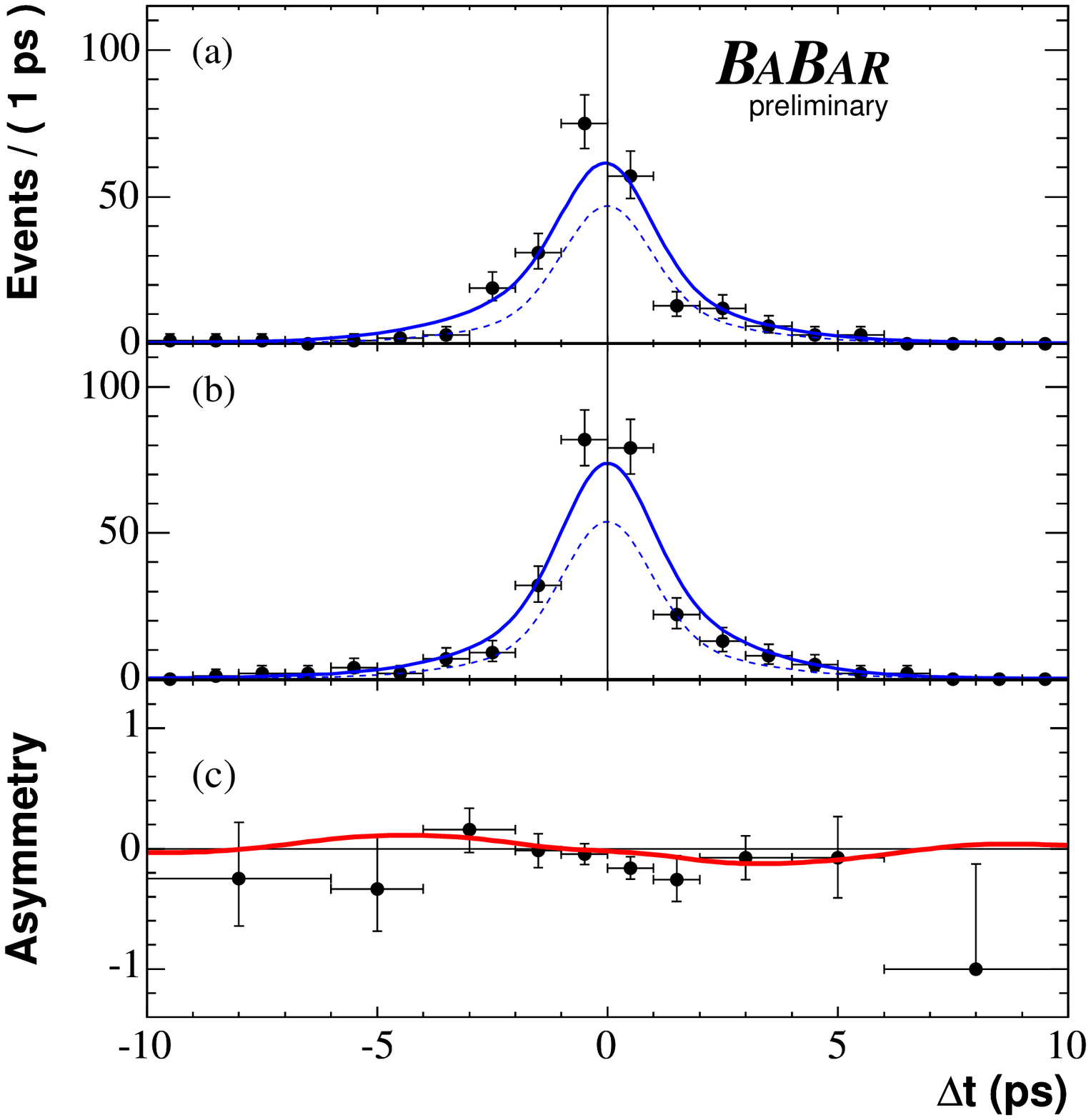}
\end{center}
  \vspace*{-0.5cm}
 \caption{Projections onto $\Delta t$ for \etapKzl\ 
(sum of the sub-decay modes \fetaprrgkl\ and \fetapreppkl )
of a subset of the data for which the signal likelihood 
exceeds a mode-dependent threshold 
that optimizes the sensitivity. Points with errors 
represent the data, solide curve the full fit functions, 
and dashed blue curve the total background functions, 
for (a) \Bz\ and (b) \Bzb\ tagged events. 
The asymmetry between \Bz\ and \Bzb\ tags is shown in (c).}
  \label{fig:DeltaTProj}
\end{figure}

The agreement between PDF simulated events and data is investigated 
using likelihood ratios. 
We generate signal and background Monte Carlo samples  from the PDFs and, 
using the fitted parameter values from nominal fit, we calculate the 
likelihoods for both samples.  
In Fig.~\ref{fig:ProbRatioDistri} we show the likelihood ratio 
$\calL(Sg)/[\calL(Sg)+\calL(Bg)]$ for
the two sub-decays \fetaprrgkl\ and \fetapreppkl\  for data and for the 
PDF generated events.
In Fig.~\ref{fig:DEplot} we show the  projection onto \DE\ of a subset of 
the data for which the signal likelihood 
(computed without the plotted variable) exceeds a mode-dependent 
threshold that optimizes the sensitivity.
We show in Fig.~\ref{fig:DeltaTProj} the $\Delta t$
projection and asymmetry for 
\BetapKzl . 

\section{Systematic Uncertainties and Cross-checks}
\label{sec:Systematics}

The contributions to the systematic uncertainties in $S$ and $C$ for
\etapKzl\ are summarized in 
Table~\ref{tab:systtab}.

We evaluate the uncertainties associated with the PDF shapes by variation of
the parameters describing each discriminating variable. Systematic errors 
associated with signal parameters 
(\deltat\ resolution function, tagging fractions, and dilutions) are determined 
by varying their values within their errors.
Uncertainties due to \deltamd\ and $\tau_B$ are obtained 
by varying these parameters by the uncertainty in their
world average values \cite{PDG2004}. All changes are combined in quadrature
obtaining an error of 0.02 for $S$ and 0.01 for $C$.

We vary the SVT alignment parameters in the signal Monte Carlo events by the size of 
misalignments found in the real data, and assign the resulting shift in the fit 
results as the systematic error of 0.01 for both $S$ and $C$.

The systematic errors due to interference between the 
CKM-suppressed $\bar{b} \rightarrow \bar{u} c \bar{d}$ amplitude and the 
favored $b \rightarrow  c \bar{u}d$
for some tag-side $B$ decays are found to be negligible for $S$ and gives a 
contribution to the $C$ uncertainty of about 0.01. 

The systematic error due to \BB\ background is estimated to be 
0.03 in $S$ and 0.01 in  $C$ parameter.
An uncertainty of 0.01 is assigned to account for limitations of Monte Carlo 
statistics and modeling of the signal.  We assign an uncertainty of 0.01
to account for the uncertainty in the position and size of the beam
spot, determined from variation of these quantities in signal MC.  
The total systematic error is obtained by summing individual errors in quadrature.

\begin{table}[htbp]
\caption{Estimates of systematic errors for  \etapKzl .}
\label{tab:systtab}
\begin{center}
\begin{tabular}{lcc}
\hline\hline
Source of error &  $\sigma(S)$ & $\sigma(C)$  \\
\hline
PDF Shapes              &$0.02$  &$0.01$     \\
SVT alignment           &$0.01$  &$0.01$    \\
Tag-side interference   &$0.00$  &$0.01$ \\
\BB\ Background         &$0.03$  &$0.01$    \\
MC statistics/modeling  &$0.01$  &$0.01$    \\
Beam spot               &$0.01$  &$0.01$    \\
\hline
Total                   &$0.04$  &$0.03$    \\
\hline\hline
\end{tabular}
\end{center}
\end{table}

We have also performed a number of checks of our results. 
When we fit the combined sample 
\etapKzs\ and  \etapKzl\ with the value for $C$
fixed to zero, we find $S= 0.37\pm0.13$.  We produce samples of
pseudo-experiments generated with events produced to match the PDF
distributions.  From these samples, we verify that the
fit bias on $S$ and $C$  is negligible and that there is a 
good agreement between expected and observed errors. 
The fit was also verified with our $\Bz \ra J/\psi K^0_L$ data sample.

\section{Conclusion}
In a sample of 232 million \BB\ pairs we have reconstructed  
\mbox{$137\pm 22$}  \fetapreppkl\ events
and   \mbox{$303\pm 49$}  \fetaprrgkl\ events.
We use these events to measure
the time-dependent asymmetry parameters $S$ and $C$:
$$
\begin{array}{ccccc}
S &=&  0.60 \pm 0.31 \,\,\,{\rm(stat)} \pm 0.04 \,\,\,{\rm(syst)}   \\
C &=&  0.10 \pm 0.21 \,\,\,{\rm(stat)} \pm 0.03 \,\,\,{\rm(syst)}   \\
\end{array}
$$

Using this sample  and  the $\etapr K^0_S$ sample found  in 
Ref.~\cite{Previous}, we obtain a total
of  $1245\pm 67$ \etapKz\ events and with a combined fit of all  data 
we measure:

$$
\begin{array}{ccccc}
S  &=& \msp0.36  \pm  0.13 \,\,\,{\rm(stat)} \pm 0.03 \,\,\,{\rm(syst)} \\
C  &=&  -0.16 \pm 0.09 \,\,\,{\rm(stat)} \pm 0.02 \,\,\,{\rm(syst)}   \\
\end{array}
$$

All these results are preliminary. Our result for $S$ differs from 
the \babar\ value of $\sin 2\beta = 0.722 \pm 0.040 \pm 0.023$
in charmonium decays~\cite{s2b} by 2.8 standard deviation.

\section{Acknowledgments}
\label{sec:Acknowledgments}

We are grateful for the 
extraordinary contributions of our \pep2\ colleagues in
achieving the excellent luminosity and machine conditions
that have made this work possible.
The success of this project also relies critically on the 
expertise and dedication of the computing organizations that 
support \babar.
The collaborating institutions wish to thank 
SLAC for its support and the kind hospitality extended to them. 
This work is supported by the
US Department of Energy
and National Science Foundation, the
Natural Sciences and Engineering Research Council (Canada),
Institute of High Energy Physics (China), the
Commissariat \`a l'Energie Atomique and
Institut National de Physique Nucl\'eaire et de Physique des Particules
(France), the
Bundesministerium f\"ur Bildung und Forschung and
Deutsche Forschungsgemeinschaft
(Germany), the
Istituto Nazionale di Fisica Nucleare (Italy),
the Foundation for Fundamental Research on Matter (The Netherlands),
the Research Council of Norway, the
Ministry of Science and Technology of the Russian Federation, and the
Particle Physics and Astronomy Research Council (United Kingdom). 
Individuals have received support from 
CONACyT (Mexico),
the A. P. Sloan Foundation, 
the Research Corporation,
and the Alexander von Humboldt Foundation.


\begin{thebibliography}{99}
\bibitem{Penguin}
Y. Grossman and  M. P. Worah, \plb{395}, 241 (1997)
\bibitem{soni}
D. Atwood and A. Soni, \plb{405}, 150 (1997)

\bibitem{lonsoni}
D.~London and A. Soni, \plb{407}, 61 (1997).

\bibitem{beneke} M. Beneke and M. Neubert, \npb{651}, 225 (2003).

\bibitem{Gross} Y. Grossman \etal, \jprd{68}, 015004 (2003).
\bibitem{Gronau}C.-W. Chiang, M. Gronau and J. L. Rosner, \jprd{68}, 074012 (2003).

\bibitem{Gronau2} M. Gronau, J. L. Rosner and J. Zupan, \plb{596}, 107 (2004);
M. Gronau, Nucl. Phys. Proc. Suppl. {\bf 142}, 263-270 (2005).


\bibitem{Isosca} \babar\ Collaboration, B. Aubert \etal, \jprl{93}, 181806 (2004)
\bibitem{PRD} \babar\ Collaboration, B. Aubert \etal, \jprd{70}, 032006 (2004)

\bibitem{BN}
M. Beneke and M. Neubert, \npb{675}, 333 (2003).
\bibitem{Previous} 
\babar\ Collaboration, B. Aubert \etal, \jprl{94}, 191802 (2005).
\bibitem{BELLE}
Belle Collaboration, K. Abe \etal, \jprl{91}, 261602 (2003).
\bibitem{babar2}
\babar\ Collaboration, B.\ Aubert \etal, \nima{479}, 1 (2002).

\bibitem{pep}
\pep2\  Conceptual Design Report, SLAC-R-418 (1993).
\bibitem{PDG2004}
Particle Data Group, S. Eidelman \etal, \plb{592}, 1 (2004).
\bibitem{s2b} 
\babar\ Collaboration, B. Aubert \etal, Phys. Rev. Lett. {\bf 94}, 161803 (2005).

\end{thebibliography}
\end{document}